\documentclass[final]{raa06}           %  For uploading to arXiv and final 
\usepackage{graphicx,times}             %for PS/EPS graphics inclusion, new
\usepackage{natbib}
\usepackage{amssymb,amsmath}
\bibpunct{(}{)}{;}{a}{}{,}
\usepackage[colorlinks=true, citecolor=blue]{hyperref}%
\usepackage{graphicx,kantlipsum,setspace}
\usepackage{caption}
\usepackage{graphics,epsf}
\usepackage{amsmath}                % American Mathematical Society package
\usepackage{amsfonts}               % American Mathematical Society fonts
\usepackage{amssymb}                % American Mathematical Society symbol
\usepackage{epsfig}                 % EPS figures
\usepackage{appendix}
\usepackage{graphicx}
\usepackage{float}
\usepackage{color}
\usepackage{multirow}
\usepackage{colortbl}
\usepackage[para,online,flushleft]{threeparttable}
\usepackage{xcolor}

\hypersetup{citecolor=blue, % color for \cite{...} links
            linkcolor=red, % color for \ref{...} links
            menucolor=blue, % color for Acrobat menu buttons
            urlcolor=blue}  % color for \url{...} links
\newcommand{\km}{{~\rm km}}
\newcommand{\s}{{~\rm s}}

\newcommand{\erg}{{~\rm erg}}

\begin{document}
\titlerunning{Difficulties of violent merger to account for type Ia Supernova}
   \title{Difficulties of two exploding white dwarfs to account for type Ia supernovae with bimodal nebular emission profiles
%\footnotetext{$*$ Supported by the National Natural Science Foundation of China.}
}
%   \subtitle{I. Place Your Subtitle Here}

   \volnopage{Vol.0 (20xx) No.0, 000--000}      %%preserved for Editor. DOn't remove!
   \setcounter{page}{1}          %%starting page, preserved for Editor. DOn't remove!

%\author[0000-0003-0375-8987]{Noam Soker}
%\author[0000-0002-9444-9460]{Dmitry Shishkin}

   \author{Jessica Braudo, Noam Soker
     % \inst{1}
    }

   \institute{Department of Physics, Technion - Israel Institute of Technology, Haifa, 3200003, Israel;   {\it    jessicab@campus.technion.ac.il; soker@physics.technion.ac.il} \\ 
   \vs\no
   {\small Jessica Jessica Braudo: orcid: {0009-0001-4877-1125}; Noam Soker: orcid: {0000-0003-0375-8987}}\\
%        \and
%             Full institute address for the third author\\
\vs\no
   {\small Received~~20xx month day; accepted~~20xx~~month day}}

\abstract{
We use a simple dynamical scheme to simulate the ejecta of type Ia supernova (SN Ia) scenarios with two exploding white dwarfs (WDs) and find that the velocity distribution of the ejecta has difficulties accounting for bimodal emission line profiles with a large separation between the two emission peaks. The essence of the dynamical code is in including the fact that the ejecta does not leave the system instantaneously. We find that the final separation velocity between the centers of masses of the two WDs' ejecta is $\simeq 80 \%$ of the pre-explosion WDs' orbital velocity, i.e., we find separation velocities of $4200-5400 \km \s^{-1}$ for two WDs of masses $M_1=M_2=0.94 M_\odot$. The lower separation velocities we find challenge scenarios with two exploding WDs to explain bimodal emission line profiles with observed velocity separations of up to $\simeq 7000 \km \s^{-1}$. Only the mass in the ejecta of one WD with an explosion velocity lower than the separation velocity contributes to one peak of the bimodal profile; this is the inner ejecta. We find the inner ejecta to be only $\lesssim 15\%$ of the ejecta mass in energetic explosions. Less energetic explosions yield higher inner mass but lower separation velocities. We encourage searching for alternative explanations of bimodal line profiles. 
 }

\maketitle

% ==========================================================
\section{INTRODUCTION}
\label{sec:intro}
% ==========================================================

There is no consensus on the leading scenarios for type Ia supernovae (SNe Ia), nor is there a consensus on how to classify the different scenarios. These disagreements are evident in hundreds of papers in recent years and in more than a dozen of reviews in the last decade  (\citealt{Maozetal2014, MaedaTerada2016, Hoeflich2017, LivioMazzali2018, Soker2018Rev, Soker2019Rev, Soker2024Rev, Wang2018,  Jhaetal2019NatAs, RuizLapuente2019, Ruiter2020, Aleoetal2023, Liuetal2023Rev, Vinkoetal2023, RuiterSeitenzahl2025}).
All SN Ia scenarios have drawbacks, some very severe regarding normal SNe Ia; still, these scenarios might account for some peculiar SNe Ia. Some observations are challenge to most (e.g., \citealt{Pearsonetal2024}), or even all (e.g., \citealt{Wangetal2024}) scenarios, such as the relation between the $\gamma$-ray escape time and the mass of ${56}Ni$ (e.g., \citealt{SchinasiLembergKushnir2024, sharonKushnir2024, Sharonetal2024}). 
For these reasons, researchers of SN Ia scenarios should be humble and remember the other scenarios besides those they are in love with. Also, studies should not freeze out in the previous millennium by citing only the single degenerate (SD) and double degenerate (DD) scenarios, but rather consider all SN Ia scenarios, including new channels of these scenarios, e.g., the common-envelope wind model \citep{MengPodsiadlowski2017, CuiMeng2022}, the carbon–oxygen–neon white dwarf (WD) with a main sequence companion channel of the SD scenario (e.g., \citealt{GuoMeng2025RAA}), and new mass transfer prescriptions (e.g., \citealt{Lietal2023RAA}).   
In Table \ref{Tab:Table1}, we present one recent classification from \cite{Soker2024Rev}, where references and more details can be found (see also \citealt{BraudoSoker2024}). 
 Since the table has been presented before, we bring it in this section. It is important to present the table to emphasize that all scenarios must be considered; none should be ignored. 
% TTTTTTTTTTTTTTTTTTTTTTTTTTTTTTTTTTTTTTTTTTTTT
\begin{table*}
%\tiny
\scriptsize
%\footnotesize
\begin{center}
  \caption{An SN Ia scenario classification}
    \begin{tabular}{| p{1.7cm} | p{1.2cm} | p{1.2cm}| p{1.8cm}| p{1.8cm} | p{1.8cm} | p{1.7cm} | p{1.8cm} |}
\hline  % ----------------------------
\textbf{Group} & \multicolumn{2}{c|}{$N_{\rm exp}=1$: {{Lonely WD}}}  &  \multicolumn{5}{c|}{$N_{\rm exp}=2$}     \\  
\hline  % ----------------------------
\textbf{Outcome} & \multicolumn{2}{c|}{Mostly normal SNe Ia}  &  \multicolumn{5}{c|}{Mostly peculiar SNe Ia}\\ 
\hline  % ----------------------------
\textbf{{SN Ia Scenario}}  & {Core Degenerate }    & {Double Degenerate - MED} & {Double Degenerate - violent merger} & {Double Detonation } & {Triple and quadruple detonation} & {Single Degenerate} & {WD-WD collision} \\
\hline  % ----------------------------
\textbf{{Name}} & CD & DD-MED & DD & DDet & DDet+ & SD-MED or SD & WWC\\
\hline  % ----------------------------
\textbf{MED time} & MED  & MED  & 0  & 0 & 0  & MED or 0 & 0 \\
\hline  % ----------------------------
 {$\mathbf{[N_{\rm sur}, M, Ej]}$$^{[{\rm 2}]}$}
  & $[0,M_{\rm Ch},{\rm S}]$ 
  & $[0,M_{\rm Ch}, {\rm S}]$ 
  & $[0,$sub-$M_{\rm Ch},{\rm N}]$
  & $[1,$sub-$M_{\rm Ch},{\rm N}]$
  & $[0,$sub-$M_{\rm Ch},{\rm N}]$
  & $[1,M_{\rm Ch},{\rm S~or~N}]$  
  & $[0,$sub-$M_{\rm Ch},{\rm N}]$ \\
\hline  % ----------------------------
     \end{tabular}
  \label{Tab:Table1}\\
\end{center}

\begin{flushleft}
\small 
Notes: An SN Ia scenarios classification scheme from \cite{Soker2024Rev}. 
In this study, we consider the violent merger channel of the DD scenario and the DDet+ scenario, where both white dwarfs (WDs) explode, resulting in channels with zero median delay (MED) time. We keep the DD abbreviation for the DD-violent merger channel as is common in many studies.   
\newline
 Abbreviation. MED time: Merger to explosion delay time (includes mass transfer to explosion delay time).    
$N_{\rm exp}$ is the number of stars in the system at the time of the explosion. $N_{\rm sur}=1$ if a companion survives the explosion while $N_{\rm sur}=0$ if no companion survives the explosion; in some peculiar SNe Ia, the exploding WD is not destroyed, and it also leaves a remnant, i.e., $N_{\rm sur}=2$. $M_{\rm Ch}$ for near-Chandrasekhar-mass explosion; sub-$M_{\rm Ch}$ for sub-Chandrasekhar mass explosion. The ejecta morphology Ej: S indicates
scenarios that can lead to a spherical SNR, while N indicates scenarios that cannot form a spherical SNR.
\end{flushleft}
\end{table*}
% TTTTTTTTTTTTTTTTTTTTTTTTTTTTTTTTTTTTTTTTTTTTT

In the double detonation (DDet) scenario, igniting an outer helium layer of the mass-accreting WD excites a shock wave that propagates into the CO WD and detonates it. Table \ref{Tab:Table1}, for example, presents the view that the DDet scenario mainly ends with peculiar SNe Ia. For example, some runaway WDs might result from type Iax SNe (a peculiar SN Ia class) rather than the DDet scenario (e.g., \citealt{Igoshevetal2023}).   
The DDet has been the subject of many studies in recent years (e.g., just from 2024, \citealt{Callanetal2024, MoranFraileetal2024, PadillaGonzalezetal2024, Polinetal2024, Rajaveletal2024, Shenetal2024, Zingaleetal2024}), but so are the other SN Ia scenarios (e.g., \citealt{Boraetal2024, Bregmanetal2024, DerKacyetal2024, Itoetal2024, Koetal2024, Kobashietal2024, Limetal2024, Palicioetal2024, Phillipsetal2024, Soker2024RAAPN, Uchidaetal2024, OHoraetal2025}).

The present study deals with scenarios where both WDs explode, e.g., the violent merger channel of the DD scenario and the triple-detonation channel of the double detonation scenario, where, in addition to the explosion of the mass-accreting WD, the mass-donor WD also explodes and leaves no remnant. This is the triple-detonation channel where the mass-donor WD that explodes is a helium WD (e.g., \citealt{Papishetal2015, Bossetal2024}), and its quadruple-detonation sub-channel where the mass-donor is a HeCO hybrid WD that explodes (e.g., \citealt{Tanikawaetal2019, Pakmoretal2022}; we mark by DDet+ the triple-detonation and quadruple-detonation, which leave no surviving stellar remnant. 
\cite{Pakmoretal2021} simulate a case where the helium detonation on the mass-accretor WD fails to ignite this WD, but ignites the mass-donor WD; the later leaves a surviving WD and is not relevant to the present study; we also do not deal with merger that lead to no explosion, e.g., \citealt{Yangetal2022}).  

In calculating the runaway velocity of the surviving WD in the DDet scenario, one must consider that the explosion does not occur instantaneously; rather, it takes several seconds for the inner ejecta layers, those that are slower, to pass the orbit of the surviving WD. During that time, the slower ejecta decelerates the surviving WD. \cite{BraudoSoker2024} simple calculations show that the deceleration phase lasts for $\simeq 10- 20 \s$, and can slow down the surviving WD by $\simeq 10 \%$. Hydrodynamical simulations of the DDet that include all effects, like of \cite{Glanzetal2024}, give the correct runaway velocity. \cite{Glanzetal2024} consider the outcome of their DDet simulation, where the explosion occurs after the mass-donor is partially stripped, to be a peculiar faint SN Ia.  Note that the inner ejecta is rich in iron group elements because it comprises the inner parts of the exploding WDs. 

Here, we consider the effect of the slower ejecta on the final relative velocity of the centers of mass of the two exploding WDs. We describe our calculations in Section \ref{sec:Numerical}, and the results in Section \ref{sec:Results}.
We summarize this short study in Section \ref{sec:Summary}.

% ===========================================
\section{The numerical procedure}
\label{sec:Numerical}
% ===========================================

We use a MATLAB code that we wrote (see \citealt{BraudoSoker2024}) that simulates the post-explosion evolution considering only gravitational forces, i.e., neglecting thermal and radiation pressures. 
Each simulation begins with the simultaneous explosion of both WDs, which have an initial mass of $M_1 = M_2 = M_{\rm WD}=0.94M_{\odot}$ and radii of $R_1 = R_2 = 0.0091R_{\odot}$ (WD radius according to \citealt{Bedard2020}). 
We also consider three explosion energies, with total (both WDs combined) energies of $E_{\rm exp} = 10^{51}\erg,  1.5 \times 10^{51}\erg$ and $2 \times 10^{51}\erg$.

Since we do not have thermal and radiation pressure, we take each WD ejecta to be at its terminal expansion velocity and homologous (velocity proportional to the distance from the center of explosion). We take the explosion ejecta of each WD to be composed of 65 numerical expanding shells as in the model of \cite{Gronowetal2021}. We calculate the mass in each numerical shell of the SN ejecta by scaling to our WD mass the density profile from \cite{Gronowetal2021} and the Heidelberg Supernova Model Archive (\textsc{HESMA}\footnote{https://hesma.h-its.org}, e.g., \citealt{Kromeretal2017}). 
The distance between the WDs at explosion is $a_{\rm ex} = (R_1+R_2)/2$, as the model of \cite{Tucker2024} that we examine in this study. Namely, the two WDs are in the process of merging at explosion.

In reality, when the ejecta of one WD reaches the center of mass of the ejecta of the second WD, the ejecta has not yet reached its terminal velocity because, at early stages, the ejecta remains hot, and not all of its thermal energy has been converted into kinetic energy. To account for this with our numerical scheme that has no thermal pressure, we also examine cases with the same ejecta model but assume that the velocity of each of the 65 numerical mass shells at our numerical explosion time, $v_{\rm s,explosion}$, is a fraction 
\begin{equation}
  \beta \equiv v_{\rm s,explosion} / v_{\rm s, terminal}
\label{eq:beta}
\end{equation}
of its terminal velocity $v_{\rm s, terminal}$. We simulate cases with $\beta = 0.85,0.9,0.95, 1$. 
Overall, we simulated 12 cases, with three explosion energies and four expansion velocity fractions $\beta$.

The orbital velocity of each of the exploding WDs around the center of mass of the binary system just before the explosion is 
\begin{equation}
    v_{\rm orb,s} = \sqrt{ 
    \frac{G M_{\rm WD}}{2 a_{\rm ex}}  
    }    .
\label{eq:VorbS}
\end{equation}

The numerical code calculates the velocity of the numerical mass shells of the ejecta of each WD over time as follows. It takes each numerical shell to maintain a spherical structure. It calculates the orbital motion of the numerical shell of one WD ejecta under the gravity of the ejecta shells of the other WD that haven’t yet crossed it. Namely, as expanding ejecta shell number $i=1,2, \ldots, 65$, $S_{1,i}$, of WD1 engulfs the center of mass of the ejecta of WD2, this shell does not influence anymore the velocity of the ejecta of WD2. Only the numerical ejecta shells of WD1 that haven’t yet crossed the center of mass of the ejecta of WD2 affect its motion gravitationally.

We take the time step to be $\Delta t = 0.001 \s$ for $0 \s < t < 2 \s$, $\Delta t = 0.01 \s$ for $2 \s < t < 4 \s$ and $\Delta t = 1 \s$ for $t > 4 \s$. We tested $\Delta t = 0.001 \s$ values for the entire simulation and observed only a 0.5\% change in the final velocity of the ejecta. We also tested $\Delta t = 0.01 \s$ values for the entire simulation and observed only a 0.3189\% change in the final velocity of the WD.

% ==============================================
\section{Results}
\label{sec:Results}
% ==============================================

We emphasize that we examine the bimodal emission profile that results from the orbital motion of the two pre-explosion WDs, as \cite{Tucker2024} claims. We do not refer to an asymmetrical explosion in the violent merger that results from the asymmetrical explosion itself rather than the orbital velocity (e.g., \citealt{Pakmoretal2012}).  

As discussed in \cite{BraudoSoker2024}, the basic dynamical process is that the ejecta requires time to engulf the other WD or its ejecta if it explodes. Namely, the orbital velocity cannot be neglected relative to the ejecta velocity, and the ejecta does not expand instantaneously. This implies that the center of mass of each ejecta shell does not maintain its pre-explosion velocity, which is the pre-explosion WD orbital velocity because the ejecta of the other WD exerts a gravitational field on it. 

We define the \textit{inner ejecta} as the ejecta mass that did not engulf (did not cross) the center of mass of the inner ejecta of the other WD. The inner ejecta with an expansion velocity lower than the relative velocity of the two centers of masses of the two ejecta, $v_{\rm sep}$, will never engulf the center of mass of the other ejecta. At late time, therefore, each shell $i$ of the inner ejecta has an expansion velocity relative to its center of mass that is $v_{{\rm in},i} < v_{\rm sep}$. We define the velocity of the inner ejecta center of mass relative to the center of mass of the WD binary system as $v_{\rm RA}$, i.e., its run-away velocity; because we have equal-mass WDs, $2 v_{\rm RA} = v_{\rm sep}$. 
We also define the ratio of the run-away velocity of the WD ejecta to the WD velocity around the center of mass at the explosion,  
\begin{equation}
    f_v \equiv \frac{v_{\rm RA}}{v_{\rm orb,s}}   .
\label{eq:fv}
\end{equation}

\begin{figure}
\begin{center}
\includegraphics[trim=0.5cm 1.0cm 0.0cm 0.0cm ,clip, width=0.51\textwidth]{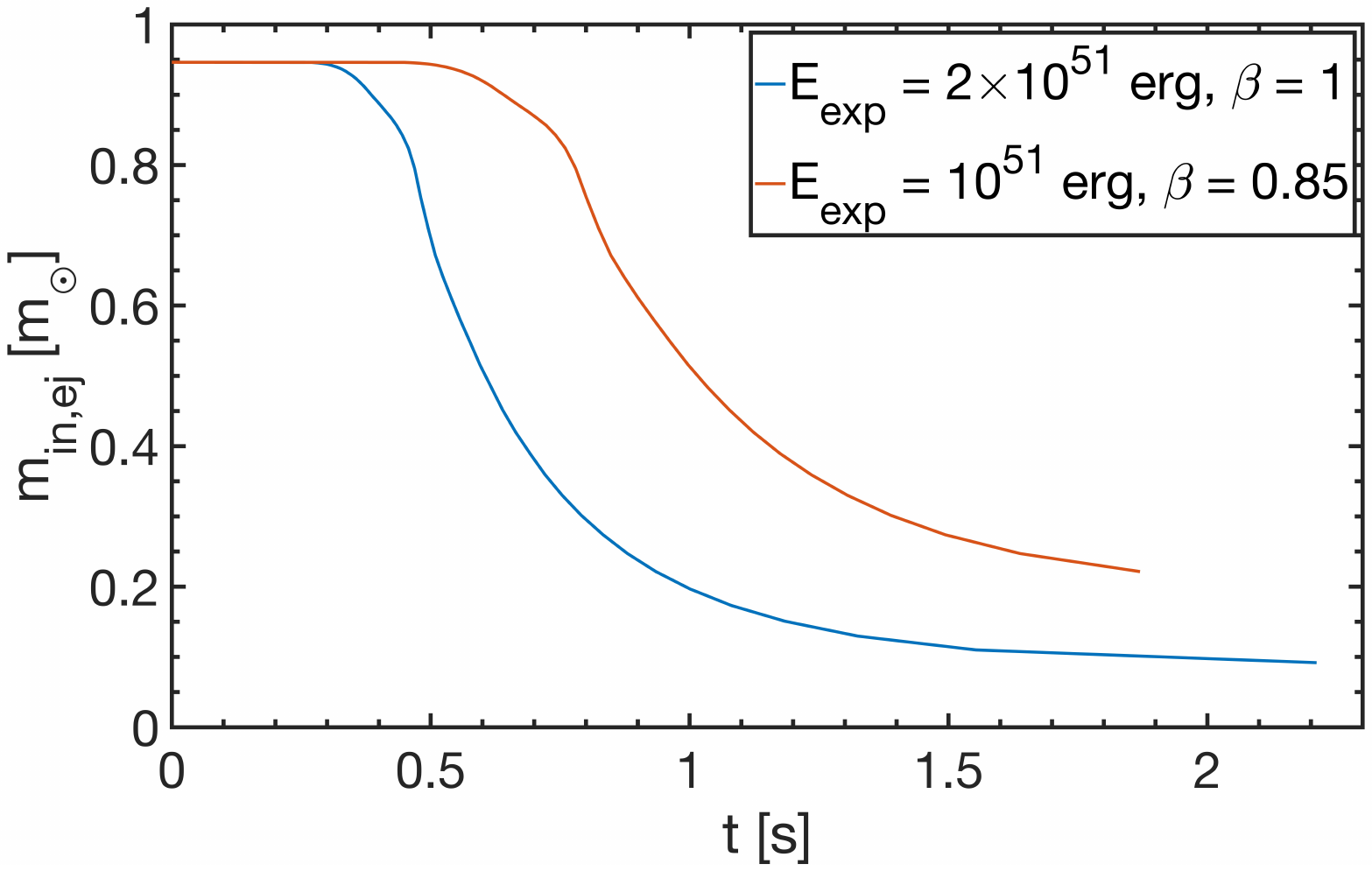} % mass_loss.pdf
% scale=0.06
\caption{The inner ejecta mass of one WD as a function of time. The inner ejecta is the mass of the shells that did not engulf (cross) the center of mass of the inner ejecta of the other WD. The blue (lower) line shows the inner ejecta of the most energetic case with total (the two WDs) explosion energy of $E_{\rm exp} = 2\times10^{51} \erg$, and $\beta = 1$ (equation \ref{eq:beta}), and the red (upper) line shows the inner mass of the least energetic case in which $E_{\rm exp} = 10^{51} \erg$ and $\beta=0.85$.  
}
\label{Fig:massloss}
\end{center}
\end{figure}

In Figure \ref{Fig:massloss}, we plot the inner mass (of one WD) as a function of time for two cases as indicated. 
It takes the ejecta about two seconds to reach its terminal mass. This is shorter than the orbital period just before the explosion, $P_{\rm orb}=6.32 \s$, but not negligible, again, implying that the ejecta is not lost instantaneously. We list the terminal inner mass for the 12 simulated cases in the fourth column of Table \ref{Tab:Table2}.
% FFFFFFFFFFFFFFFFFFFFFFFFFFFFFFFFFF

% FFFFFFFFFFFFFFFFFFFFFFFFFFFFFFFFFF
%
% TTTTTTTTTTTTTTTTTTTTTTTTTTTTTTTT
\begin{table}
%\tiny
\scriptsize
%\footnotesize
\begin{center}
  \caption{Terminal velocity and inner mass}
    \begin{tabular}{| p{1.2cm} | p{0.8cm} | p{1.0cm}| p{1.0cm}| p{1.0cm} | p{1.0cm} |}
\hline  % ----------------------------
{${E_{\rm exp}}$} & $\beta$ & {${f_{\rm v}}$} & {${m_{\rm in, ej,t}}$} & {${m_{\rm c,t} }$} \\
$[10^{51} \erg]$  &         &                 & $[m_{\odot}]$ & $[m_\odot]$ \\
\hline  % ----------------------------
1 & 0.85 & 0.697 & 0.220 & 0.011\\
\hline  % ----------------------------
1 & 0.9 & 0.729 & 0.195 & 0.008\\
\hline  % ----------------------------
1 & 0.95 & 0.755 & 0.195 & 0.006\\
\hline  % ----------------------------
1 & 1 & 0.763 & 0.172 & 0.006\\
\hline  % ----------------------------
1.5 & 0.85 & 0.784 & 0.149 & 0.012\\
\hline  % ----------------------------
$1.5$ & 0.9 & 0.793 & 0.149 & 0.012\\
\hline  % ----------------------------
1.5 & 0.95 & 0.816 & 0.128 & 0.008\\
\hline  % ----------------------------
1.5 & 1 & 0.822 & 0.128 & 0.008\\
\hline  % ----------------------------
2 & 0.85 & 0.820 & 0.128 & 0.017\\
\hline  % ----------------------------
2 & 0.9 & 0.839 & 0.109 & 0.012\\
\hline  % ----------------------------
2 & 0.95 & 0.846 & 0.109 & 0.012\\
\hline  % ----------------------------
2 & 1 & 0.864 & 0.091 & 0.008\\     
\hline  % ----------------------------
\end{tabular}
  \\
\label{Tab:Table2}
\end{center}
\begin{flushleft}
\small 
Notes: ${E_{\rm exp}}$ is the combined total explosion energy of both WDs. ${\beta}$ is the fraction of the terminal expansion velocity of each expanding numerical mass shell. ${E_{\rm exp}}$ and $\beta$ are the parameters of each simulation. ${f_{\rm v}}$ is the ratio of the final velocity of the center of mass of the inner ejecta to the orbital velocity of the WD at the explosion $v_{\rm WD}=3148 \km \s^{-1}$. ${m_{\rm in, ej,t}}$ is the terminal inner ejecta mass of each WD, meaning it is the mass of the numerical shells that did not engulf the center of the other inner ejecta. ${m_{\rm c,t} }$ is the mass of the inner ejecta of one WD that collides with the other inner ejecta. The total colliding mass is twice this mass for two WDs. 
our non-hydrodynamical simulation does not follow the collision, 
\end{flushleft}
\end{table}
% TTTTTTTTTTTTTTTTTTTTTTTTTTTTTTTTTTTTTTTTTTT

\begin{figure}
\begin{center}
	% \centering
%	\hspace*{-2cm} 
	% This cut edges: [trim=left bottom right top, clip]{file}
%	\hspace{1cm}
\includegraphics[trim=6.0cm 2.0cm 0.0cm 2.0cm ,clip, scale=0.50]{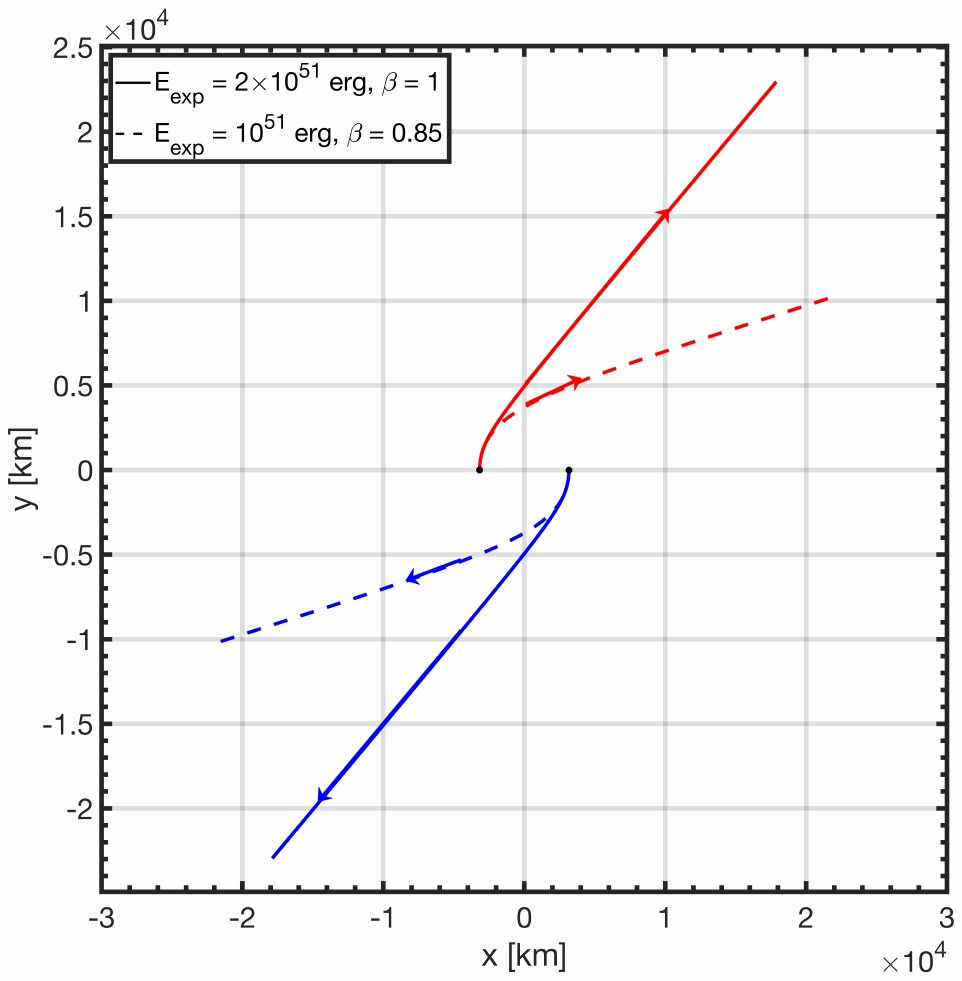} % trajectory_both.pdf
% scale=0.06
\caption{The trajectories of the two centers of mass of the two ejecta masses from the two exploding WDs, from explosion to $t= 10 \s$ for two cases. The black dots are the location of the two WDs at explosion ($t=0$). Solid lines show the trajectory of the most energetic case in which $E_{\rm exp} = 2\times10^{51} \erg$ and $\beta = 1$, and dashed lines show the trajectory of the least energetic case in which $E_{\rm exp} = 10^{51} \erg$ and $\beta = 0.85$. Each color stands for one WD. The WDs' ejecta is rotating around each other right after the explosion. At explosion, the left WD moves in the $+y$ direction, and the right WD moves in the $-y$ direction. 
}
\label{Fig:trajectory}
\end{center}
\end{figure}

Because the ejecta that escapes (the mass that is not the inner ejecta) is not lost instantaneously, and there is always an inner mass that exerts gravity on the other WD ejecta, the center of mass terminal velocity of the inner ejecta relative to the center of mass of the binary system is lower than the pre-explosion orbital velocity of the WD, and at a different direction. In our non-hydrodynamical simulations, we take the ejecta mass that exerts gravity on the other ejecta to be only the mass that did not engulf (did not cross) the center of mass of the other ejecta, i.e., the inner ejecta (Section \ref{sec:Numerical}).
In Figure \ref{Fig:trajectory}, we show the trajectory of the centers of masses of the inner ejecta of the two WDs for the cases with $E_{\rm exp} =  2 \times 10^{51} \erg$ and $\beta=1$ (solid lines) which yield the maximum terminal velocity (Table \ref{Tab:Table2}), and $E_{\rm exp} =  1 \times 10^{51} \erg$ and $\beta=0.85$ (dashed lines) which yield the slowest terminal velocity in the cases we simulate. 
Because the two WDs are of equal mass, the two trajectories in each case are opposite.  At explosion ($t=0$), one WD moves in the $+y$ and the other in the $-y$ direction. After the explosion, the center of mass trajectory bends tens of degrees (different values for the 12 cases). This figure further emphasizes that the orbital velocity of the WD at explosion is not the terminal velocity of the center of mass of the ejecta.  
% FFFFFFFFFFFFFFFFFFFFFFFFFFFFFFFFFFFFFFFFFFFFFFF

% FFFFFFFFFFFFFFFFFFFFFFFFFFFFFFFFFFFFFFFFFFFFFFF

\begin{figure*}[]
	\begin{center}
%	\hspace*{-2cm} 
	% This cut edges: [trim=left bottom right top, clip]{file}
%	\hspace{1cm}
\includegraphics[trim=0.0cm 2.0cm 0.0cm 1.0cm ,clip, scale=0.63]{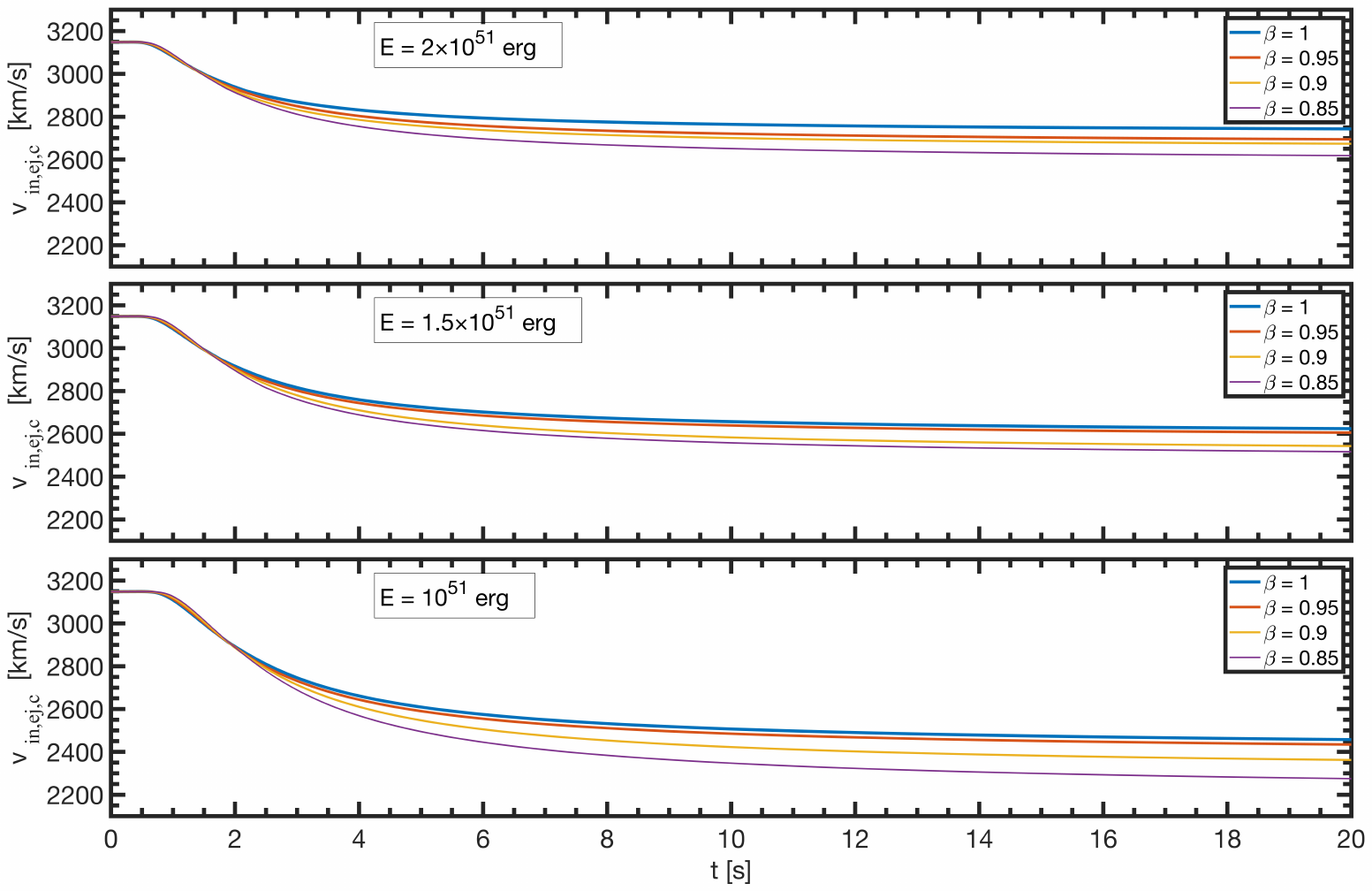}  % velocities.pdf
\caption{The velocity of the center of mass of the inner ejecta of one WD relative to the center of mass of the binary system as a function of time.  Each panel shows a case with a different explosion energy according to the insets. In all cases, the masses of the two WDs are $M_1=M_2=0.94 M_\odot$. The colored lines are for different values of $\beta$ (equation \ref{eq:beta}).  
}
%\vskip+0.5cm
\label{Fig:Velocities}
\end{center}
\end{figure*}

This study examines the claim that the bimodal emission profile of SNe Ia results from the explosion of two WDs. \cite{Tucker2024} consider the emission of the two peaks to result from two WD ejecta with expansion velocity relative to its center of mass that is smaller than the velocity between the two WD ejecta $v_{\rm sep}$. We, therefore, examine the velocity of the center of mass of the inner ejecta (the ejecta that did not cross the center of mass of the other inner ejecta). In Figure \ref{Fig:Velocities}, we present the velocity of the center of mass of the inner ejecta of one WD relative to the center of mass of the binary system (the other WD ejecta is symmetrical, i.e., the same speed in the opposite direction).    
% FFFFFFFFFFFFFFFFFFFFFFFFFFFFFFFFFFFFFFFFFF
% FFFFFFFFFFFFFFFFFFFFFFFFFFFFFFFFFFFFFFFFFff

Figure \ref{Fig:Velocities} and the third column of Table \ref{Tab:Table2} show that the terminal velocity of the center of mass of the inner ejecta is much lower than the orbital velocity of the two WDs at the explosion, $f_v=0.697 - 0.864$ for the cases we simulate. The cases that give faster velocities, those in the lower rows of Table \ref{Tab:Table2}, leave very low mass in the inner ejecta. Only the inner ejecta emission contributes to an emission peak separated from the other WD ejecta's emission peak.  

% ==========================================
\section{Discussion and Summary}
\label{sec:Summary}
% ==========================================

We performed a simple dynamical, i.e., non-hydrodynamical, calculation of the ejecta of two equal-mass WDs, $M_1=M_2=0.94 M_\odot$, that explode in the violent merger channel of the DD scenario of SNe Ia. We found that the mass of the ejecta of a WD that stays separated from the other ejecta, i.e., the inner ejecta, is a small fraction of the WD mass; Figure \ref{Fig:massloss} for evolution with time, and the fourth column of Table \ref{Tab:Table2} for the terminal inner ejecta mass. Because the ejecta does not leave the explosion site instantaneously, the two ejecta masses of the two WDs do not maintain their orbital velocity at the explosion. The velocity of the center of mass of the ejecta relative to the explosion site changes direction after the explosion (Figure \ref{Fig:trajectory}) and substantially decreases (Figure \ref{Fig:Velocities} and the third column of Table \ref{Tab:Table2}).

\cite{Tucker2024} argue that the separate ejecta masses of the two exploded WDs in the violent merger channel explain some of SNe Ias' bimodal nebular emission profiles. The bimodal profile is one where a spectral line comprises two peaks at different wavelengths. \cite{Tucker2024} attributes the two peaks to two separate parcels of ejecta resulting from two exploded WDs. 

A SN Ia is classified as bimodal if the two velocity components are separated
by more than their combined width. This implies that only the inner ejecta contributes to the bimodal line profile. \cite{Vallelyetal2020} convolved an ejecta velocity profile with a template of SN Ia spectrum to obtain bimodal emission lines (following \citealt{Dongetal2015}). The velocity distribution (the Kernel) often requires the two ejecta masses to be completely separated in the velocity profile. We do not find such velocity profiles as, in our cases, the ejecta of the two WDs collide with each other.  

\cite{Tucker2024} calculate that for a velocity separation between the two velocity components of $v_{\rm sep} \gtrsim 6000 \km \s^{-1}$ the combined mass of the exploding WDs is $\gtrsim 1.8 M_\odot$. This was our motivation to take $M_1=M_2=0.94 M_\odot$. 
However, we find that for reasonable explosion energies, the final separation velocity is much lower than that of the two WDs at the explosion. The fastest case (the lowest row in Table \ref{Tab:Table2}) gives $v_{\rm sep} = 2 v_{\rm RA} = 2 f_v v_{\rm orb,s} = 5440 \km \s^{-1}$. However, this case requires a total explosion energy of $E_{\rm exp} = 2 \times 10^{51} \erg$, which is on the high side of SNe Ia, and, more importantly, only a small fraction of the initial mass, $ <15 \%$, stays in the inner ejecta that contributes to the bimodal profile. The rest of the ejecta mass of both WDs expand at higher velocities than $v_{\rm sep}$ and form a broad emission line, i.e., wider than the separation between the two ejecta.   
Less energetic explosions yield more inner ejecta mass, but they have much lower $v_{\rm sep} =2 f_v v_{\rm orb,s}$ and are far from the observed $v_{\rm sep} \gtrsim 6000 \km \s^{-1}$ in some bimodal SNe Ia (Figure 4 in \citealt{Tucker2024}). 

We find another potential problem with the violent merger explanation of bimodal SNe Ia. If the emission profile is due to two ejecta clumps, then the total width of the two lines (the red-shifted and blue-shifted ejecta clumps) should be broader than that of an SN Ia with no bimodal profile. This is not the situation in all cases, e.g., compare the bimodal SN Ia with the non-bimodal SN Ia that \cite{Tucker2024} presents in Figure 1 of his paper. This requires further studies and an enlarged sample of SNe Ia.  

\cite{Tucker2024} also mentions the WWC scenario  (see Table \ref{Tab:Table1}) as a possible explanation for bimodal SNe Ia, as earlier studies suggested, e.g., \cite{Dongetal2015}. However, the WWC scenario is extremely rare (e.g.,  section 2 in the first astro-ph version of \citealt{Sokeretal2014}; \citealt{Toonenetal2018, Michaelyetal2021}), and we consider it unlikely to explain bimodal SNe Ia. 

We suggest two solutions to the difficulty we identified.  The first is that the iron group elements in the scenarios where two WDs explode are concentrated at lower velocities than some simulations indicate. This would give the required velocity profiles with separate Fe/Co/Ni ejecta masses. However, this will not solve the difficulty of achieving the highest separation velocities of $v_{\rm sep} \simeq 6000-7000 \km \s^{-1}$.   
The second possibility is a Chandrasekhar mass WD explosion that ejects an `iron bullet.’ The Tycho supernova remnant has such an iron clump (e.g., \citealt{Yamaguchietal2017}). \citealt{TsebrenkoSoker2015} suggest that the iron clump is the remnant of an `iron bullet’ ejected at a somewhat higher velocity than the ejecta. It is unclear what forms this iron bullet, but it is observed, and its velocity separates it from the inner iron group ejecta. 

We note that some effects we did not include in the simple toy model will likely strengthen our claims. Some simulations (e.g., \citealt{Pakmoretal2022, Bossetal2024} show that the ejecta from the WD that explodes first wraps around the other WD; the second WD explodes with material around it, making the explosion very complicated (e.g, \citealt{Peretsetal2019, Zenatietal2023}). The flow of the first WD material around the other will smear the formation of two different ejecta, weakening the bimodal emission. Other effects might also play a role, like magnetic fields of the mass accreting WD (e.g., \citealt{CuiLi2022RAA}).

The main result of this study's calculations is that the ejecta of SNe Ia of the violent merger channel, or more generally of channels where two WDs explode, cannot be considered to leave the system instantaneously. Namely, the pre-explosion orbital velocity cannot be neglected as it is not much lower than the ejecta velocity, particularly of the inner layer of the exploding WDs. The final separation velocity of the two ejecta masses of the two WDs is much slower than the WDs' orbital velocity at the explosion. From our main calculations' result, we challenge the claim of  \cite{Tucker2024} that the violent merger channel of the DD scenario best explains bimodal SNe Ia. We encourage searching for alternative explanations, such as an iron bullet.

% ===============================
\section*{Acknowledgments}
% ===============================

We thank Michael Tucker for his useful comments and clarifications, as well as an anonymous referee for their support.  

%%%%%%%%%%%%%%%%%%%%%%%%%%%

%%%%%%%%%%%%%%%%%%%%%%%%%%%

% %%%%%%%%%%%%  References %%%%%%%%%%%%%%%%%%%%%


\newpage 

\begin{thebibliography}{}

\bibitem[\protect\citeauthoryear{Aleo et al.}{2023}]{Aleoetal2023} Aleo P.~D., Malanchev K., Sharief S., Jones D.~O., Narayan G., Foley R.~J., Villar V.~A., et al., 2023, ApJS, 266, 9. %doi:10.3847/1538-4365/acbfba

\bibitem[\protect\citeauthoryear{B{\'e}dard et al.}{2020}]{Bedard2020} B{\'e}dard A., Bergeron P., Brassard P., Fontaine G., 2020, ApJ, 901, 93. doi:10.3847/1538-4357/abafbe

\bibitem[\protect\citeauthoryear{Boos, Townsley, \& Shen}{2024}]{Bossetal2024} Boos S.~J., Townsley D.~M., Shen K.~J., 2024, ApJ, 972, 200. %doi:10.3847/1538-4357/ad5da2

\bibitem[\protect\citeauthoryear{Bora et al.}{2024}]{Boraetal2024} Bora Z., K{\"o}nyves-T{\'o}th R., Vink{\'o} J., B{\'a}nhidi D., B{\'\i}r{\'o} I.~B., Bostroem K.~A., B{\'o}di A., et al., 2024, PASP, 136, 094201. %doi:10.1088/1538-3873/ad6e18

\bibitem[\protect\citeauthoryear{Braudo \& Soker}{2024}]{BraudoSoker2024} Braudo J., Soker N., 2024, OJAp, 7, 7. %doi:10.21105/astro.2310.16554

\bibitem[\protect\citeauthoryear{Bregman et al.}{2024}]{Bregmanetal2024} Bregman J.~N., Gnedin O.~Y., Seitzer P.~O., Qu Z., 2024, ApJL, 968, L6. %doi:10.3847/2041-8213/ad498f

\bibitem[\protect\citeauthoryear{Callan et al.}{2024}]{Callanetal2024} Callan F.~P., Collins C.~E., Sim S.~A., Shingles L.~J., Pakmor R., Srivastav S., Pollin J.~M., et al., 2024, arXiv, arXiv:2408.03048. %doi:10.48550/arXiv.2408.03048

\bibitem[\protect\citeauthoryear{Cui et al.}{2022}]{CuiMeng2022} Cui Y., Meng X., Podsiadlowski P., Song R., 2022, A\&A, 667, A154. %doi:10.1051/0004-6361/202141335

\bibitem[\protect\citeauthoryear{Cui \& Li}{2022}]{CuiLi2022RAA} Cui Z., Li X.-D., 2022, RAA, 22, 025001. %doi:10.1088/1674-4527/ac3744


\bibitem[\protect\citeauthoryear{DerKacy et al.}{2024}]{DerKacyetal2024} DerKacy J.~M., Ashall C., Hoeflich P., Baron E., Shahbandeh M., Shappee B.~J., Andrews J., et al., 2024, ApJ, 961, 187. %doi:10.3847/1538-4357/ad0b7b

\bibitem[\protect\citeauthoryear{Dong et al.}{2015}]{Dongetal2015} Dong S., Katz B., Kushnir D., Prieto J.~L., 2015, MNRAS, 454, L61. %doi:10.1093/mnrasl/slv129

\bibitem[\protect\citeauthoryear{Glanz et al.}{2024}]{Glanzetal2024} Glanz H., Perets H.~B., Bhat A., Pakmor R., 2024, arXiv, arXiv:2410.17306. %doi:10.48550/arXiv.2410.17306


\bibitem[\protect\citeauthoryear{Gronow et al.}{2021}]{Gronowetal2021} Gronow S., Collins C.~E., Sim S.~A., R{\"o}pke F.~K., 2021, A\&A, 649, A155. %doi:10.1051/0004-6361/202039954

\bibitem[\protect\citeauthoryear{Guo et al.}{2025}]{GuoMeng2025RAA} Guo B., Meng X., Tian Z., Luo J., Liu Z., 2025, RAA, 25, 015018. %doi:10.1088/1674-4527/ada2e9

\bibitem[\protect\citeauthoryear{Hoeflich}{2017}]{Hoeflich2017} Hoeflich P., 2017, in Handbook of Supernovae, Springer International Publishing AG, 2017, p. 1151 %doi:10.1007/978-3-319-21846-5\_56


\bibitem[\protect\citeauthoryear{Igoshev, Perets, \& Hallakoun}{2023}]{Igoshevetal2023} Igoshev A.~P., Perets H., Hallakoun N., 2023, MNRAS, 518, 6223. %doi:10.1093/mnras/stac3488


\bibitem[\protect\citeauthoryear{Ito et al.}{2024}]{Itoetal2024} Ito D., Sano H., Nakazawa K., Mitsuishi I., Fukui Y., Sudou H., Takaba H., 2024, arXiv, arXiv:2405.11285. %doi:10.48550/arXiv.2405.11285


\bibitem[Jha et al.(2019)]{Jhaetal2019NatAs} Jha, S.~W., Maguire, K., \& Sullivan, M.\ 2019, Nature Astronomy, 3, 706

\bibitem[\protect\citeauthoryear{Ko et al.}{2024}]{Koetal2024} Ko T., Suzuki H., Kashiyama K., Uchida H., Tanaka T., Tsuna D., Fujisawa K., et al., 2024, ApJ, 969, 116. %doi:10.3847/1538-4357/ad4d99

\bibitem[\protect\citeauthoryear{Kobashi et al.}{2024}]{Kobashietal2024} Kobashi R., Lee S.-H., Tanaka T., Maeda K., 2024, ApJ, 961, 32. %doi:10.3847/1538-4357/ad05c2

\bibitem[\protect\citeauthoryear{Kromer, Ohlmann, \& R{\"o}pke}{2017}]{Kromeretal2017} Kromer M., Ohlmann S., R{\"o}pke F.~K., 2017, MmSAI, 88, 312. %%doi:10.48550/arXiv.1706.09879

\bibitem[\protect\citeauthoryear{Li, Liu, \& Wang}{2023}]{Lietal2023RAA} Li L.-H., Liu D.-D., Wang B., 2023, RAA, 23, 075010. %doi:10.1088/1674-4527/acd0ea

\bibitem[\protect\citeauthoryear{Lim et al.}{2024}]{Limetal2024} Lim G., Im M., Paek G.~S.~H., Yoon S.-C., Imsng Team, 2024, ASPC, 536, 29

\bibitem[\protect\citeauthoryear{Liu, R{\"o}pke, \& Han}{2023}]{Liuetal2023Rev} Liu Z.-W., R{\"o}pke F.~K., Han Z., 2023, RAA, 23, 082001. %doi:10.1088/1674-4527/acd89e


\bibitem[Livio \& Mazzali(2018)]{LivioMazzali2018} Livio, M., \& Mazzali, P.\ 2018, Physics Reports, 736, 1 


\bibitem[Maeda, \& Terada(2016)]{MaedaTerada2016} Maeda, K., \& Terada, Y.\ 2016, International Journal of Modern Physics D, 25, 1630024

\bibitem[Maoz et al.(2014)]{Maozetal2014} Maoz, D., Mannucci, F., \& Nelemans, G.\ 2014, \araa, 52, 107

\bibitem[\protect\citeauthoryear{Meng \& Podsiadlowski}{2017}]{MengPodsiadlowski2017} Meng X., Podsiadlowski P., 2017, MNRAS, 469, 4763. %doi:10.1093/mnras/stx1137

\bibitem[\protect\citeauthoryear{Michaely}{2021}]{Michaelyetal2021} Michaely E., 2021, MNRAS, 500, 5543. %doi:10.1093/mnras/staa3623

\bibitem[\protect\citeauthoryear{Mor{\'a}n-Fraile et al.}{2024}]{MoranFraileetal2024} Mor{\'a}n-Fraile J., Holas A., R{\"o}pke F.~K., Pakmor R., Schneider F.~R.~N., 2024, A\&A, 683, A44. %doi:10.1051/0004-6361/202347769

\bibitem[\protect\citeauthoryear{O'Hora et al.}{2024}]{OHoraetal2025} O'Hora J., Ashall C., Shahbandeh M., Hsiao E., Hoeflich P., Stritzinger M.~D., Galbany L., et al., 2024, arXiv, arXiv:2412.09352

\bibitem[\protect\citeauthoryear{Padilla Gonzalez et al.}{2024}]{PadillaGonzalezetal2024} Padilla Gonzalez E., Howell D.~A., Terreran G., McCully C., Newsome M., Burke J., Farah J., et al., 2024, ApJ, 964, 196. %doi:10.3847/1538-4357/ad19c9

\bibitem[\protect\citeauthoryear{Pakmor et al.}{2022}]{Pakmoretal2022} Pakmor R., Callan F.~P., Collins C.~E., de Mink S.~E., Holas A., Kerzendorf W.~E., Kromer M., et al., 2022, MNRAS, 517, 5260. %doi:10.1093/mnras/stac3107

\bibitem[\protect\citeauthoryear{Pakmor et al.}{2012}]{Pakmoretal2012} Pakmor R., Kromer M., Taubenberger S., Sim S.~A., R{\"o}pke F.~K., Hillebrandt W., 2012, ApJL, 747, L10. doi:10.1088/2041-8205/747/1/L10

\bibitem[\protect\citeauthoryear{Pakmor et al.}{2021}]{Pakmoretal2021} Pakmor R., Zenati Y., Perets H.~B., Toonen S., 2021, MNRAS, 503, 4734. 
%doi:10.1093/mnras/stab686


\bibitem[\protect\citeauthoryear{Palicio et al.}{2024}]{Palicioetal2024} Palicio P.~A., Matteucci F., Della Valle M., Spitoni E., 2024, A\&A, 689, A203. %doi:10.1051/0004-6361/202449740


\bibitem[\protect\citeauthoryear{Papish et al.}{2015}]{Papishetal2015} Papish O., Soker N., Garc{\'\i}a-Berro E., Aznar-Sigu{\'a}n G., 2015, MNRAS, 449, 942. %doi:10.1093/mnras/stv337

\bibitem[\protect\citeauthoryear{Pearson et al.}{2024}]{Pearsonetal2024} Pearson J., Sand D.~J., Lundqvist P., Galbany L., Andrews J.~E., Bostroem K.~A., Dong Y., et al., 2024, ApJ, 960, 29. %doi:10.3847/1538-4357/ad0153

\bibitem[\protect\citeauthoryear{Perets et al.}{2019}]{Peretsetal2019} Perets H.~B., Zenati Y., Toonen S., Bobrick A., 2019, arXiv, arXiv:1910.07532. %doi:10.48550/arXiv.1910.07532

\bibitem[\protect\citeauthoryear{Phillips et al.}{2024}]{Phillipsetal2024} Phillips M.~M., Ashall C., Brown P.~J., Galbany L., Tucker M.~A., Burns C.~R., Contreras C., et al., 2024, ApJS, 273, 16. doi:10.3847/1538-4365/ad4f7e

\bibitem[\protect\citeauthoryear{Pollin et al.}{2024}]{Polinetal2024} Pollin J.~M., Sim S.~A., Pakmor R., Callan F.~P., Collins C.~E., Shingles L.~J., R{\"o}pke F.~K., et al., 2024, MNRAS, 533, 3036. %doi:10.1093/mnras/stae1909

\bibitem[\protect\citeauthoryear{Rajavel, Townsley, \& Shen}{2024}]{Rajaveletal2024} Rajavel N., Townsley D.~M., Shen K.~J., 2024, arXiv, arXiv:2408.10981. %doi:10.48550/arXiv.2408.10981

\bibitem[\protect\citeauthoryear{Ruiter}{2020}]{Ruiter2020} Ruiter A.~J., 2020, IAUS, 357, 1. %doi:10.1017/S1743921320000587

\bibitem[\protect\citeauthoryear{Ruiter \&  Seitenzahl}{2025}]{RuiterSeitenzahl2025} Ruiter A.~J.,  Seitenzahl, I.~R. 2025,
arXiv:2412.01766

\bibitem[Ruiz-Lapuente(2019)]{RuizLapuente2019} Ruiz-Lapuente, P.\ 2019, \nar, 85, 101523

\bibitem[\protect\citeauthoryear{Schinasi-Lemberg \& Kushnir}{2024}]{SchinasiLembergKushnir2024} Schinasi-Lemberg E., Kushnir D., 2024, arXiv, arXiv:2410.02849. %doi:10.48550/arXiv.2410.02849

\bibitem[\protect\citeauthoryear{Sharon \& Kushnir}{2024}]{sharonKushnir2024} Sharon A., Kushnir D., 2024, arXiv, arXiv:2407.06859. %doi:10.48550/arXiv.2407.06859

\bibitem[\protect\citeauthoryear{Sharon, Kushnir, \& Schinasi-Lemberg}{2024}]{Sharonetal2024} Sharon A., Kushnir D., Schinasi-Lemberg E., 2024, arXiv, arXiv:2407.07417. %doi:10.48550/arXiv.2407.07417

\bibitem[\protect\citeauthoryear{Shen, Boos, \& Townsley}{2024}]{Shenetal2024} Shen K.~J., Boos S.~J., Townsley D.~M., 2024, arXiv, arXiv:2405.19417. %doi:10.48550/arXiv.2405.19417

\bibitem[\protect\citeauthoryear{Soker}{2018}]{Soker2018Rev} Soker N., 2018, SCPMA, 61, 49502. %doi:10.1007/s11433-017-9144-4

\bibitem[\protect\citeauthoryear{Soker}{2019a}]{Soker2019Rev} Soker N., 2019a, NewAR, 87, 101535. %doi:10.1016/j.newar.2020.101535


\bibitem[\protect\citeauthoryear{Soker}{2024}]{Soker2024Rev} Soker N., 2024, OJAp, 7, 31. %doi:10.33232/001c.117147

\bibitem[\protect\citeauthoryear{Soker}{2024}]{Soker2024RAAPN} Soker N., 2024, RAA, 24, 015012. %doi:10.1088/1674-4527/ad0ded

\bibitem[\protect\citeauthoryear{Soker}{2024}]{Soker2024Comment} Soker N., 2024, arXiv, arXiv:2406.07363. %doi:10.48550/arXiv.2406.07363


\bibitem[\protect\citeauthoryear{Soker, Garc{\'\i}a-Berro, \& Althaus}{2014}]{Sokeretal2014} Soker N., Garc{\'\i}a-Berro E., Althaus L.~G., 2014, MNRAS, 437, L66. %doi:10.1093/mnrasl/slt142

\bibitem[\protect\citeauthoryear{Tanikawa et al.}{2019}]{Tanikawaetal2019} Tanikawa A., Nomoto K., Nakasato N., Maeda K., 2019, ApJ, 885, 103. %doi:10.3847/1538-4357/ab46b6

\bibitem[\protect\citeauthoryear{Toonen, Perets, \& Hamers}{2018}]{Toonenetal2018} Toonen S., Perets H.~B., Hamers A.~S., 2018, A\&A, 610, A22. %doi:10.1051/0004-6361/201731874
\bibitem[\protect\citeauthoryear{Tsebrenko \& Soker}{2015}]{TsebrenkoSoker2015} Tsebrenko D., Soker N., 2015, MNRAS, 453, 166. %doi:10.1093/mnras/stv1641



\bibitem[\protect\citeauthoryear{Tucker}{2024}]{Tucker2024} Tucker M.~A., 2024, arXiv, arXiv:2408.00840. %doi:10.48550/arXiv.2408.00840


\bibitem[\protect\citeauthoryear{Uchida et al.}{2024}]{Uchidaetal2024} Uchida H., Kasuga T., Maeda K., Lee S.-H., Tanaka T., Bamba A., 2024, ApJ, 962, 159. %doi:10.3847/1538-4357/ad1ff3

\bibitem[\protect\citeauthoryear{Vallely et al.}{2020}]{Vallelyetal2020} Vallely P.~J., Tucker M.~A., Shappee B.~J., Brown J.~S., Stanek K.~Z., Kochanek C.~S., 2020, MNRAS, 492, 3553. %doi:10.1093/mnras/staa003

\bibitem[\protect\citeauthoryear{Vink{\'o}, Szalai, \& K{\"o}nyves-T{\'o}th}{2023}]{Vinkoetal2023} Vink{\'o} J., Szalai T., K{\"o}nyves-T{\'o}th R., 2023, Univ, 9, 244. %doi:10.3390/universe9060244

\bibitem[\protect\citeauthoryear{Wang}{2018}]{Wang2018} Wang B., 2018, RAA, 18, 049. %doi:10.1088/1674-4527/18/5/49

  
\bibitem[\protect\citeauthoryear{Wang et al.}{2024}]{Wangetal2024} Wang Q., Rest A., Dimitriadis G., Ridden-Harper R., Siebert M.~R., Magee M., Angus C.~R., et al., 2024, ApJ, 962, 17. %doi:10.3847/1538-4357/ad0edb

\bibitem[\protect\citeauthoryear{Yamaguchi et al.}{2017}]{Yamaguchietal2017} Yamaguchi H., Hughes J.~P., Badenes C., Bravo E., Seitenzahl I.~R., Mart{\'\i}nez-Rodr{\'\i}guez H., Park S., et al., 2017, ApJ, 834, 124. %doi:10.3847/1538-4357/834/2/124

\bibitem[\protect\citeauthoryear{Yang, Thomas Tam, \& Yang}{2022}]{Yangetal2022} Yang H.-W., Thomas Tam P.-H., Yang L., 2022, RAA, 22, 105014. %doi:10.1088/1674-4527/ac8d7e

\bibitem[\protect\citeauthoryear{Zenati et al.}{2023}]{Zenatietal2023} Zenati Y., Perets H.~B., Dessart L., Jacobson-Gal{\'a}n W.~V., Toonen S., Rest A., 2023, ApJ, 944, 22. %doi:10.3847/1538-4357/acaf65

\bibitem[\protect\citeauthoryear{Zingale et al.}{2024}]{Zingaleetal2024} Zingale M., Chen Z., Rasmussen M., Polin A., Katz M., Smith Clark A., Johnson E.~T., 2024, ApJ, 966, 150. %doi:10.3847/1538-4357/ad3441


\end{thebibliography}
\end{document}